%% file: arxiv.tex
\documentclass[acmsmall,nonacm]{acmart}

\usepackage{booktabs,multirow,makecell}
\usepackage{algorithm}
\usepackage{algpseudocode}

\makeatletter
\newcommand{\fs@algorithmslim}{%
  \def\@fs@cfont{\bfseries}%
  \let\@fs@capt\floatc@ruled
  \def\@fs@pre{\hrule height.8pt depth0pt}%
  \def\@fs@post{\hrule\relax}%
  \def\@fs@mid{\hrule\relax}%
  \let\@fs@iftopcapt\iftrue}
\floatstyle{algorithmslim}
\restylefloat{algorithm}
\makeatother

\begin{document}

\title{Automated Synthesis of Heterogeneous, Hierarchical, Scoped Coherence Protocols}

\author{Fletch Rydell}
\affiliation{\institution{Duke University}\city{Durham}\country{USA}}
\email{fletch.rydell@duke.edu}

\author{An Qi Zhang}
\affiliation{\institution{University of Utah}\city{Salt Lake City}\country{USA}}
\email{an.qi.zhang@utah.edu}

\author{Nicolai Oswald}
\affiliation{\institution{NVIDIA}\city{Santa Clara}\country{USA}}
\email{mail@nicolai-oswald.de}

\author{Andr\'es Goens}
\affiliation{\institution{TU Darmstadt}\city{Darmstadt}\country{Germany}}
\email{andres.goens@tu-darmstadt.de}

\author{Vijay Nagarajan}
\affiliation{\institution{University of Utah}\city{Salt Lake City}\country{USA}}
\email{vijay@cs.utah.edu}

\author{Daniel Sorin}
\affiliation{\institution{Duke University}\city{Durham}\country{USA}}
\email{sorin@ee.duke.edu}

\date{}

\input{abstract}

\maketitle

\input{intro}
\input{background}
\input{synthesis-overview}
\input{synthesis-implementation-transaction-based}
\input{case-studies}
\input{verification-short}

\input{related-work}
\input{conclusions}

\section*{Acknowledgments}
This work is supported by ARM and by the National Science Foundation under grants CCF-2525270 and CCF-2525271.
\bibliographystyle{ACM-Reference-Format}
\bibliography{refs}

\end{document}

%% file: abstract.tex
\begin{abstract}

Processor design is converging on a new model of cache-coherent shared memory characterized by heterogeneity, hierarchy, and scopes. Protocols like CXL or AMBA CHI are used as global protocols to combine multiple clusters, each with its own cluster-level coherence protocols. Manually designing shims to interface between these cluster and global protocols is subtle and error-prone. Automatic synthesis, on the other hand, makes simplifying assumptions like the single-writer multiple-reader (SWMR) invariant, that sacrifice performance for simplicity and guaranteed correctness.

We present a shim API \textemdash a generic abstraction that enables us to classify protocol transactions by their semantic coherence guarantees. Our automated synthesis engine, ShimGen, uses this shim API to automatically compose protocols with both SWMR and relaxed accesses, taking advantage of modern architectural optimizations like scoped memory accesses and lazy invalidation.
We demonstrate ShimGen's efficacy on two case studies.  First, we compare its output to a manually-designed hierarchical protocol for the AMD APU released in gem5.
ShimGen's output is similar to the existing protocol. However, we identify one scenario where the manually-designed protocol fails to uphold compound consistency, while ShimGen's does. Second, we compare the performance of a hierarchical protocol with a global protocol that exploits non-SWMR accesses to a hierarchical protocol with a strictly SWMR global protocol. The result shows the performance benefits of accommodating global protocols with non-SWMR behavior.

\end{abstract}

%% file: intro.tex
\section{Introduction}

\begin{figure}[]
\begin{center}
    
    \includegraphics[width=0.75\linewidth]{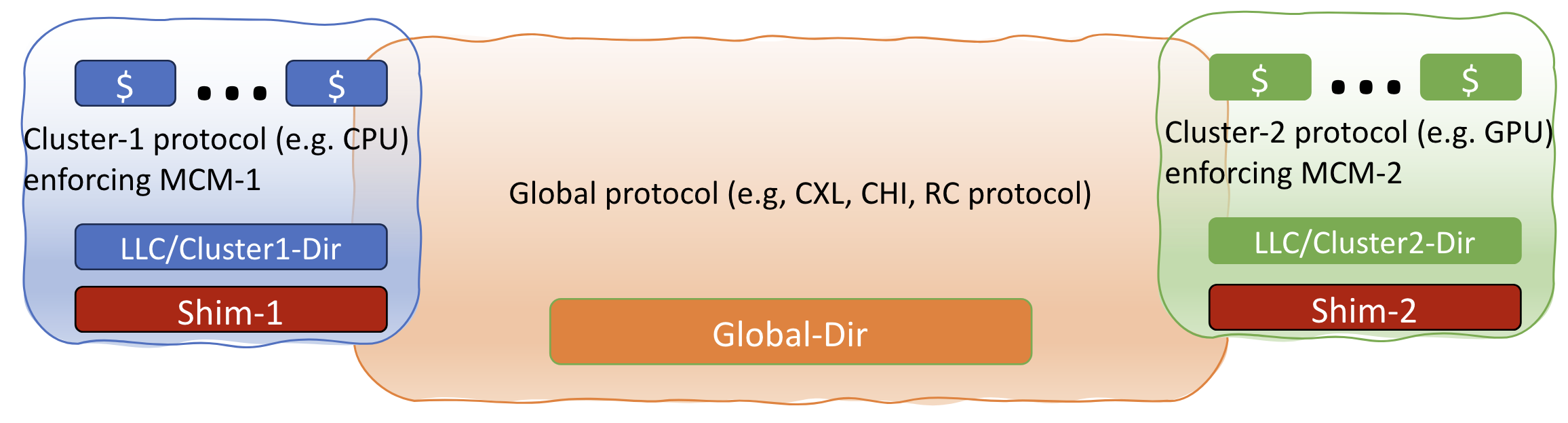}
    \caption{Heterogeneous processor made of two or more clusters---with each cluster using a distinct coherence protocol enforcing a distinct MCM---fused using a global protocol. 
    }
    \label{fig:system-model}
\end{center}
\end{figure}

Modern shared-memory systems typically integrate diverse processing elements---perhaps a cluster of ARM CPU cores alongside a cluster of NVIDIA GPU cores---onto a single die or multi-chip module or across servers, connected by a cache-coherent interconnect. These devices often enforce very different memory consistency models (MCMs) and employ very different internal coherence protocols.  They are fused through a global interconnect such as AMBA CHI~\cite{chi} or CXL~\cite{cxl}. The challenge for the architect is to design the interface logic---or $shims$---that translate coherence requests between each cluster protocol and the global protocol, as illustrated in Figure~\ref{fig:system-model}, such that each device enforces its original MCM. This correctness criterion is formalized as compound consistency~\cite{DBLP:journals/pacmpl/Goens0SAON23}.

Designing shims is difficult because the cluster and global protocols differ in fundamental ways. CPU clusters typically enforce the single-writer, multiple-reader (SWMR) invariant: a write obtains exclusive ownership, and sharers are invalidated. GPU clusters instead use self-invalidation and write-back protocols (e.g., RCC~\cite{nagarajan:primer:2020}) in which no persistent permissions are held. Global protocols like CXL support both paradigms, offering 14 distinct transaction types varying in permission, ordering, and data requirements. The space of possible translations between a cluster protocol and a global protocol is large, and the correctness criterion is subtle.

Recent work has made the first step in automating this process. vCXLGen~\cite{lefort:asplos:2026} synthesizes shims that bridge cluster protocols with CXL, producing correct protocols validated by model checking. However, vCXLGen makes a simplifying assumption: every cluster transaction is mapped to a global SWMR read or write; that is, every cluster read obtains global shared permission and every cluster write obtains global exclusive permission.

\begin{figure}[]
\begin{center}
    \includegraphics[width=0.95\linewidth]{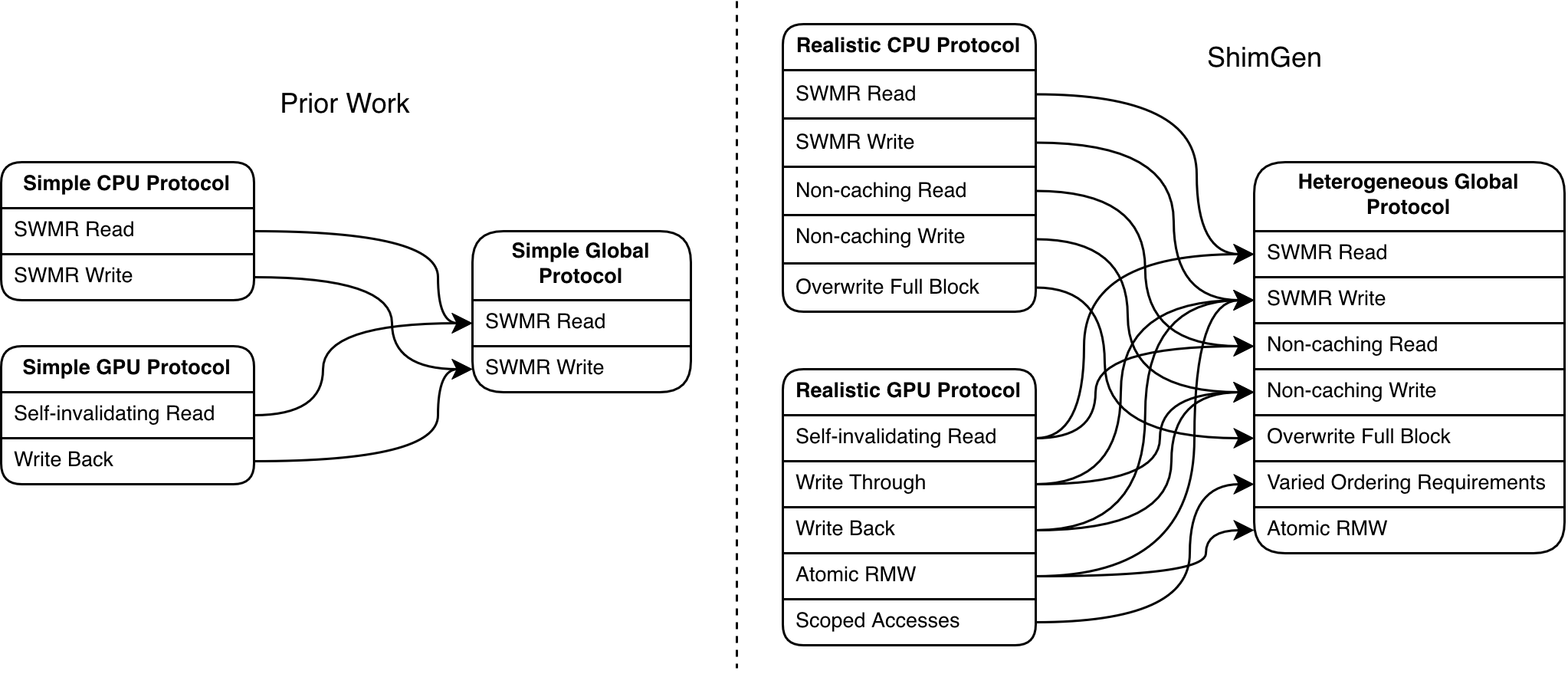}
    \caption{Comparison to prior work.
    Mapping every cluster transaction to obtain global SWMR permission (top) ignores various functionality supported by real protocols (below).}
    \label{fig:mappings-comparison}
\end{center}
\end{figure}

This simplifying assumption is insufficient for two reasons. First, it is potentially wasteful: many cluster transactions do not require global SWMR permission, and acquiring it adds unnecessary overhead. GPU protocols employ scoped consistency models in which cluster-scoped accesses need not be made globally visible at all. CPU protocols include non-temporal stores that propagate a value without requiring ownership of the block. In both cases, the cluster needs only to perform a one-off access in the global protocol---not obtain exclusive ownership---and global protocols such as CXL provide transactions precisely for this purpose (e.g., CXL's WrInv).  

Second, the assumption is restrictive: many global protocols offer a richer interface than SWMR reads and writes, and a tool that can only map to SWMR uses a fraction of what the protocol provides. Spandex~\cite{alsop:isca:2018} was proposed precisely because a pure SWMR interface is insufficient for heterogeneous systems. CXL actually provides both SWMR and non-SWMR transactions, including weakly-ordered writes and reads that do not obtain permission. Global protocols targeting multi-GPU systems, such as HMG~\cite{ren:hpca:2020}, omit SWMR permissions entirely. A synthesis tool restricted to SWMR mappings cannot exploit any of these. 

In this paper, we present ShimGen, an automated synthesis engine that addresses these limitations, as illustrated in Figure~\ref{fig:mappings-comparison}. Given state machine specifications of a cluster and global protocol, ShimGen synthesizes the shim logic to  produce a complete hierarchical protocol enforcing compound consistency.  

The key idea is a \textit{shim API}: a semantic classification of protocol transactions by the coherence guarantees they require, independent of the specific protocol messages that implement them. ShimGen classifies every transaction in the device and the global protocol into one of three categories---obtains (a cache acquires permission and/or data), one-offs (an access is performed at the directory without acquiring permission), and revokes (permission is recalled from a cluster)---with parameters capturing permission level, data requirements, and scope. For example, an MOESI GetM and a CXL RdOwn both  classify as "obtain exclusive write permission with data"; a GPU write-through  is classified as a "one-off global write." This classification decouples the semantics of a transaction from the protocol that implements it, enabling ShimGen to map between arbitrary cluster and global protocols---including non-SWMR protocols that prior tools cannot handle.  While the shim API takes inspiration from Spandex~\cite{alsop:isca:2018}, its purpose is different. Spandex defines a fixed coherence interface supporting both SWMR and non-SWMR requests. ShimGen's shim API is not itself a protocol; it is a classification used during synthesis to map any cluster protocol to any global protocol. Spandex is, in fact, one of the global protocols ShimGen can target.

We verify the correctness of ShimGen on systems composed of a wide variety of cluster and global protocols, including MOESI, CXL, CHI, and Spandex~\cite{alsop:isca:2018}, as well as extensions supporting non-temporal accesses, full-block writes, and scoped release consistency. For each system, we use model checking to verify that the resulting hierarchical protocol enforces compound consistency.

We evaluate ShimGen with two case studies. First, we compare ShimGen's output to the manually-designed AMD APU protocol in gem5~\cite{gem5:2020}, comprising a CPU MESI cluster and GPU VIPER cluster connected by a global MOESI protocol. ShimGen produces an equivalent protocol for the CPU cluster but identifies a scenario in the GPU cluster in which the manual protocol does not enforce compound consistency, whereas ShimGen's does. Second, we generate two protocol variants for a multi-cluster CXL system: one restricted to SWMR mappings (as in prior work) and one exploiting one-off writes for non-temporal stores. The latter achieves higher effective memory bandwidth on streaming workloads, revealing a concrete benefit of the richer shim API.

In summary, this work presents ShimGen, an automated synthesis engine for generating verifiably correct hierarchical, heterogeneous protocols. ShimGen takes state machine specifications of cluster and global protocols as input and produces complete shim logic enforcing compound consistency. The key contributions are:
  
\begin{itemize}
 
\item First support for non-SWMR global protocols and scopes. ShimGen is the first synthesis tool that supports global protocols without SWMR permissions, as well as the first to exploit scoped operations to avoid unnecessary global propagation, addressing the two main limitations of prior work identified above.
 
\item A shim API for heterogeneous protocols. To achieve this generality, we define a classification of protocol transactions---obtains, one-offs, and revokes---that decouples coherence semantics from protocol-specific messages. This API accommodates SWMR, non-SWMR, scoped, and hybrid protocols, including CXL and CHI, and supports protocol features (non-temporal accesses, scoped operations) that prior synthesis tools~\cite{lefort:asplos:2026, oswald:hpca:2022, oswald:isca:2020} cannot handle.
  
\item Comparison to a manually-designed protocol. We compare ShimGen's output to the AMD APU protocol in gem5, comprising a CPU MESI cluster and GPU VIPER cluster connected by a global MOESI protocol. ShimGen produces an equivalent protocol for the CPU cluster. For the GPU cluster, the two protocols differ in one scenario: the manually-designed protocol does not enforce compound consistency whereas ShimGen's does.

\end{itemize}

%% file: background.tex
\section{Background}

\subsection{Memory Consistency Models} 
\label{sec:back:mcm}

\vspace{0.05in}
\noindent{\bf Hardware MCMs.} The hardware MCM is part of the instruction set specification and specifies how loads and stores are ordered~\cite{nagarajan:primer:2020}. 
All major commercial (homogeneous) processors support precisely defined MCMs~\cite{sewell:cacm:2010,arm,waterman:risc-v:2014}.

\vspace{0.05in}
\noindent{\bf Scoped GPU MCMs.} Typically, GPUs expose the hierarchical nature of the architecture through a thread hierarchy known as scopes.
Scopes enable efficient synchronization in GPUs because they allow the synchronizing loads and stores to propagate to only  those threads involved in the synchronization.
For the purpose of this paper, 
we consider two scopes: \emph{global} and \emph{cluster}. A memory operation marked \texttt{cluster} scope means its effects propagate locally within the cluster, whereas a memory operation marked \texttt{global} means that its effects are propagated globally across all clusters.
Scopes internal to a cluster, such as CTA scope in NVIDIA GPUs, do not affect the composition of protocols.

\vspace{0.05in}
\noindent{\bf Compound MCMs.}
The advent of heterogeneous systems has inspired the definition of system-wide compound MCMs that can accommodate clusters with different intra-cluster MCMs~\cite{DBLP:journals/pacmpl/Goens0SAON23}. 
Consider the heterogeneous machine shown in Figure~\ref{fig:system-model}, where cluster-1 enforces MCM-1 and cluster-2 enforces MCM-2. This heterogeneous machine should satisfy the compound MCM. This is not a new MCM; rather it is a compositional amalgamation of MCM-1 and MCM-2. 
An x86-TSO store, for example, continues to behave like an x86-TSO store even after the x86-TSO cluster is fused with, say, a GPU. Because of this compositional property, compound MCMs retain compiler mappings. That is, if a language-level MCM has provably correct mappings to MCM M1, the mappings are retained even when the cluster enforcing M1 is fused with other clusters. 

\subsection{Coherence Protocols}
\label{sec:protocol-descriptions}
\label{interface}

A coherence protocol is an important component of the system's enforcement of the MCM.
Some coherence protocols---especially the ones targeted towards CPUs---obtain persistent read and write permissions for blocks.
Such protocols uphold a Single-Writer-Multiple-Reader (SWMR) invariant, ensuring that reads see the most up-to-date value throughout the system and written values are globally visible immediately.

SWMR protocols commonly use a subset of the MOESI states.
In MOESI protocols, a GetS(hared) message requests read permission and adds the cache to a list of block sharers.
A GetM(odified) messages requests read/write permission, causing Inv(alidation) requests to be sent to any sharers; the cache obtains write permission only after all Inv(alidation)-Ack(nowledgments) have been received from sharers.
When a cache owns a block, requests are forwarded to it so it can respond and downgrade permissions if necessary.

Not all protocols enforce SWMR.
GPU-based protocols such as RCC (release-consistency-directed coherence)~\cite{nagarajan:primer:2020} instead allow readers to read stale data, relying on self-invalidations and flushes on acquires and releases to enforce consistency.
With RCC, cache controllers only send GetV(alid) and PutV(alid) messages to an L2 controller to read and write back data when necessary---the protocol does not track coherence permissions or forward requests and invalidations.
These messages may also include information about the scope of a synchronization access, allowing cluster and global-scoped synchronization to be handled differently in a hierarchical system.
Because exclusive ownership cannot be obtained in an L1, GPU protocols also typically include dedicated messages for performing atomic read-modify-writes at the LLC.

Some protocols combine both SWMR permission-based and RCC-like permissionless reads, writes, and RMWs.
For example, Spandex~\cite{alsop:isca:2018} provides both ReqS(hared) and ReqO(wned) requests obtaining read and write permission and ReqV(alid) and ReqW(rite)T(hrough) requests reading and writing values at the LLC without obtaining permission.
CXL~\cite{cxl} similarly provides both permission-based RdShared and RdOwned requests, as well as RdCurr (a read of the current value without obtaining permission) and both strongly and weakly ordered WrInvs (a write performed at the host after pulling data from the requestor with a WritePull message).

Referring back to Figure~\ref{fig:system-model}, all these different types of protocols are employed as cluster protocols and global protocols, and they must interoperate. 
We aim to produce a tool flexible enough to handle a wide range of possible protocols.

%% file: synthesis-overview.tex
\section{ShimGen Overview}

\subsection{What ShimGen Takes as Input}
\label{shimgen-input}

ShimGen takes in specifications of the coherence protocol for each cluster involved in a system, as well as the coherence protocol for the global interconnect.
Each protocol specification consists of a finite state machine for the cache controller and directory/LLC controller, following a model similar to the protocol descriptions in Nagarajan et al.~\cite{nagarajan:primer:2020}.
These specifications are provided using a domain-specific language (DSL) based on that used in prior research on automated protocol generation~\cite{oswald:isca:2018, oswald:isca:2020, oswald:hpca:2022}.

ShimGen aims to automatically generate shims between cluster and global protocols without requiring annotations on coherence messages and transactions.
In most cases, the user is only required to label each access type supported by the protocol (e.g., load, store, acquire, and release for RCC) as a read, write, or RMW.
However, there are two cases where additional consistency information is needed.
First, if using a consistency-directed global protocol including accesses with weak ordering guarantees, ShimGen must take these consistency semantics into account when combining protocols.
To do this, the user is required to specify the translation between the cluster and global protocol accesses, generated with a tool such as ArMOR~\cite{lustig:isca:2015}.
Second, if the cluster MCM supports scopes, ShimGen can avoid unnecessarily propagating cluster-scoped operations globally on the critical path.
This requires the user to specify which accesses in the cluster protocol are cluster-scoped, rather than global scoped.
By default, unlabeled accesses are assumed to be strongly-ordered and globally-scoped.
Note that in each of these cases, the user is only required to give information about the consistency semantics of accesses, not the protocol messages or transactions used to implement them.

\subsection{What ShimGen Produces as Output}

ShimGen produces a ``shim'' between a cluster protocol and the global protocol.
This shim translates incoming messages at the cluster directory into required behavior in the global cache controller, and incoming messages at the global cache controller to behavior in the cluster directory.
ShimGen's end products are three state machines: a cluster directory modified to incorporate its associated shim logic, a global cache modified to incorporate its associated shim logic, and a single controller that fuses these two modules.

The first two outputs augment the input cluster directory and global cache controllers to include communication between the cluster and global protocol where necessary, as well as any additional state needed to track ongoing transactions in the other protocol.
For example, whereas the original cluster directory may handle a request by immediately responding, integrating the cluster in a hierarchical system may require instead requesting permission from the global protocol, moving to a ShimGen-generated transient state to wait for a response, and sending a response only when an acknowledgment is received from the global cache controller.
Using these two controllers requires implementing some means for the shim requests discussed in Section~\ref{sec:step1-classify} and associated responses to be conveyed between them.
Outputting two separate controllers with a well-defined interface between them enables the abstraction-based verification described in Section~\ref{sec:verification}.

ShimGen also combines the cluster directory and global cache controllers together to produce one fused cache/directory controller, as done in prior research on heterogeneous and hierarchical protocol generation~\cite{oswald:isca:2020, oswald:hpca:2022}.
This single controller interfaces directly with both the cluster L1s and global directory via their respective interconnection networks.
ShimGen includes a gem5-compatible SLICC backend allowing this fused controller, along with the unmodified cluster L1s and global directory, to be used in gem5.

\subsection{Assumptions and Limitations}
\label{assume}

ShimGen assumes directory-based protocols: it assumes that each of the cluster protocols and the global protocol in the input specifications includes a directory controller, which may be physically integrated with the last-level cache (LLC).

ShimGen does not currently handle bus-based snooping protocols.
The synthesis methodology does not currently support protocols that rely on timestamps for ordering or update-based protocols. While ShimGen covers a significant range of widely implemented protocols---including the MOESI family, scoped GPU-style protocols employing self-invalidation and writebacks, and hybrid protocols (combining SWMR and non-SWMR interfaces) like Spandex and CXL---it may not encompass all academic proposals for protocols. 

As we focus here on SWMR protocols and RCC variants, it might seem like we are ignoring prevalent consistency models such as x86-TSO.  However, 
it is possible to realize many MCMs using these coherence protocols. For example, many x86-TSO systems combine SWMR coherence with a post-commit write buffer.

%% file: synthesis-implementation-transaction-based.tex
\section{ShimGen Methodology}

This section explains the three-step methodology ShimGen uses to automatically synthesize hierarchical protocols consisting of a wide variety of cluster and global protocols.
Each protocol implements some coherence guarantees for caches within its domain (between L1 caches in a cluster protocol or between L2 caches in the global protocol).
Maintaining compound consistency when protocols are combined in a hierarchical system requires shims to ensure these guarantees continue to hold in the larger system.
For example, in an MSI cluster protocol, ensuring writes to an L1 in M are visible to readers throughout the system requires obtaining a similar notion of  ownership in the global protocol.
For an RCC cluster protocol, the L1 controllers ensure that writes are propagated to the L2 in an order that respects release consistency, and the shim must ensure these writes are likewise propagated to any other clusters in a way that respects RC.

\begin{figure}
    \centering
    \includegraphics[width=1.0\linewidth]{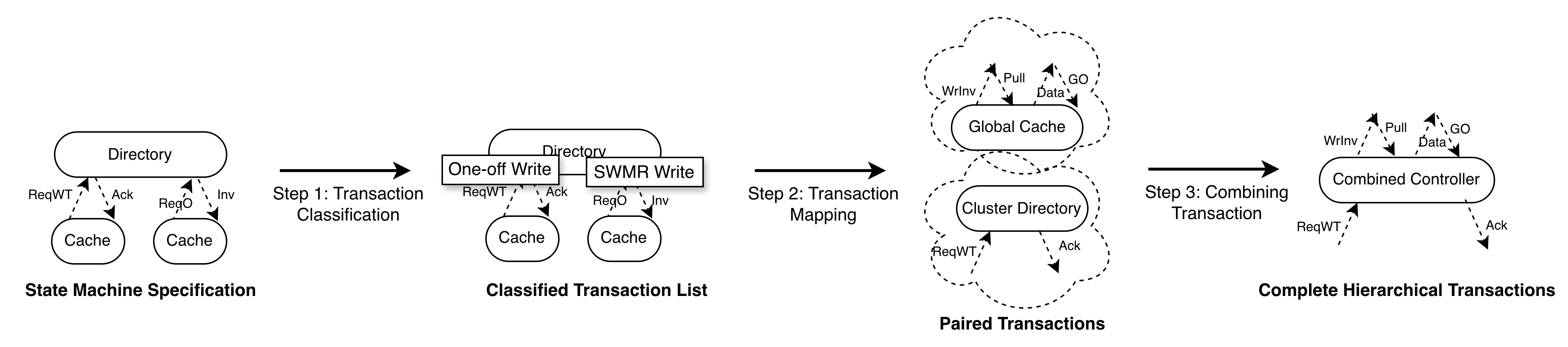}
    \caption{ShimGen Steps, Fusing a Spandex ReqWT Transaction With a CXL Global Protocol}
    \label{fig:steps}
    \vspace{-0.15in}
\end{figure}

ShimGen generates shims which uphold the expected coherence guarantees with the following three-step process, shown in Figure~\ref{fig:steps}.

\noindent \underline{Step 1: Classify transaction functionality.}
For each transaction in the cluster directory or global cache controller, identify the functionality that is needed from the rest of the system via the shim.
Because we aim to automatically fuse arbitrary protocols, we must determine this
functionality somewhat abstractly; for example, a GetM being received at an MSI cluster directory
is identified as a more generic "obtain SWMR write permission and data" transaction.
However, the choice of functionality in the shim cannot be too general,
or it may not provide the required guarantees. For example,
mapping an MSI cluster GetM to a generic “write” in the
global protocol, as done in prior work~\cite{oswald:isca:2020, oswald:hpca:2022}, will not
satisfy compound consistency if the global protocol does not
obtain exclusive ownership on writes.

To support all of the functionalities required by different protocol transactions, we developed a \textbf{shim API}: a set of actions to which ShimGen maps transactions in the cluster and global protocol.
Our shim API is based on the functionality needed by the MOESI, CXL, Spandex, and RCC protocols in Section \ref{sec:protocol-descriptions}.
These protocols encompass a wide range of coherence protocol behavior, and we have also checked that other protocols such as CHI~\cite{chi}, RCC-O~\cite{nagarajan:primer:2020}, DeNovo~\cite{choi:pact:2011}, and HMG~\cite{ren:hpca:2020} fit our API.

\noindent \underline{Step 2: Identify transactions providing this functionality.}
Given the semantics of each transaction in the cluster directory or global cache controller,
identify the transaction in the global or cluster protocol, respectively, which provides the necessary guarantees from the rest of the system.
For example, an MSI cluster GetM obtains SWMR write permission with data, so it would be mapped to a Spandex ReqO+data transaction.

In general, there is not necessarily one correct mapping between transactions upholding the required coherence guarantees.
ShimGen automatically identifies the set of correct mappings for a transaction,
allowing the architect to experiment with different possibilities,
while also selecting a reasonable default for full automation.

\noindent \underline{Step 3: Fuse cluster and global transactions.}
In the previous step, ShimGen identified the appropriate transaction in one protocol for each transaction in the other.
However, each of these transactions may include several different actions, which ShimGen must determine how to interleave.
For example, even if both the cluster and global protocol are MSI, a GetM at the cluster directory must invalidate sharers within the cluster, send data to the requestor, issue a GetM globally, collect acknowledgments from other clusters, and handle concurrent requests from the cluster and global protocols.

In this step, ShimGen determines how these operations should be interleaved to produce a correct and performant hierarchical protocol, producing a set of transitions which correctly combine the functionality of the cluster and global transaction.
By combining the transitions produced for all transactions identified in the first two steps, ShimGen produces a complete hierarchical controller.

\newcommand{\Txns}{\mathsf{Txns}}
\newcommand{\Stable}{\mathsf{Stable}}
\newcommand{\start}{\mathsf{start}}
\newcommand{\finish}{\mathsf{end}}
\newcommand{\trigger}{\mathsf{guard}}
\newcommand{\firstsend}{\mathsf{firstSend}}
\newcommand{\class}{\mathsf{class}}
\newcommand{\scope}{\mathsf{scope}}
\newcommand{\None}{\mathsf{None}}
\newcommand{\Read}{\mathsf{Read}}
\newcommand{\ReadWrite}{\mathsf{ReadWrite}}
\newcommand{\Write}{\mathsf{Write}}
\newcommand{\RMW}{\mathsf{RMW}}
\newcommand{\accesstype}{\mathsf{accType}}
\newcommand{\accessesdata}{\mathsf{accessesCacheData}}
\newcommand{\sendsdata}{\mathsf{sendsData}}

\newcommand{\reqperm}{\mathsf{reqPerm}}
\newcommand{\reqscope}{\mathsf{reqScope}}
\newcommand{\perm}{\mathsf{perm}}
\newcommand{\actions}{\mathsf{actions}}
\newcommand{\op}{\mathsf{op}}
\newcommand{\Obtain}{\mathsf{Obtain}}
\newcommand{\OneOff}{\mathsf{OneOff}}
\newcommand{\Revoke}{\mathsf{Revoke}}
\newcommand{\Global}{\mathsf{Global}}
\newcommand{\Cluster}{\mathsf{Cluster}}
\newcommand{\NoOp}{\mathsf{NoOp}}
\newcommand{\suchthat}{\mathsf{\;with\;}}

\subsection{Preliminaries}

ShimGen represents a controller as a finite state machine (FSM) graph whose vertices are per-cache-block states and edges are transitions between them.
Each transition is guarded by a condition, such as an access, incoming message, or request to evict a block, and performs a sequence of actions, such as sending messages, updating state, or reading/writing data.
A transition from state $A$ to $B$ with guard $g$ and actions $a$ is denoted by $A \xrightarrow{g / a} B$.
Transitions occur atomically, performing all actions and transitioning to a new state immediately upon receiving a message or other guard.

We define a "transaction" to be a subgraph of a controller FSM, starting with a single transition from a stable state, and including all subsequent transitions until another stable state is reached.
We write $\Txns(X)$ for the set of transactions in controller $X$, and $\Stable(X)$ for the set of stable states in $X$. For a transaction $\tau$, $\start(\tau)$ and $\trigger(\tau)$ are the start state and guard from which it begins, and $\finish(\tau)$ denotes the set of stable states it can reach.

Each cache controller input to ShimGen includes an associated set of accesses $A$ which the protocol may receive from the core.
As described in Section~\ref{shimgen-input}, each $a \in A$ is labeled with an access type $\accesstype(a)$, either $\Read$, $\Write$, or $\RMW$, and a $\scope(a)$, either cluster or global.

\subsection{Step 1: Classify Transaction Functionality}
\label{sec:step1-classify}

ShimGen's first step is to classify transactions at the cluster protocol directory or global protocol cache according to their functionality.
This step is performed independently of the protocol being fused with, analyzing the cluster protocol or global protocol individually.

We map coherence transactions to a \emph{shim API}, describing all functionality that is relevant to the construction of shims in a wide range of protocols.
This API is divided into three categories, the first two for cluster directory transactions (requiring some action in the global protocol via the shim) and the third for global cache transactions (requiring some action in an attached cluster):

\begin{itemize}
\item Obtains: Transactions granting a cache permission and/or data for a block to perform an access.
\item One-offs: Accesses performed at the directory/LLC, rather than obtaining a block with permission.
\item Revokes: Transactions in the global cache controller in which some permission is revoked or data called back.
\end{itemize}

Each of these categories includes several parameters by which a transaction is classified.
We describe the categories, their parameters, and how they are identified by analyzing the input protocols in the following subsections.
Note that ShimGen classifies transactions completely automatically, without requiring additional user inputs except those mentioned in Section \ref{shimgen-input}.
Algorithm~\ref{alg:step1} describes, in pseudocode, the classification of cluster directory transactions $\tau \in \Txns(D)$ into obtains and one-offs, which we denote $\class(\tau)$.

\subsubsection{Obtains}
\label{sec:identify-obtains}

Our first category of transactions is those that involve a cache obtaining a block with some permission so that accesses performed there are visible throughout the rest of the system.
We term such transactions ``obtains," and classify them further with two parameters: the level of permission obtained and whether up-to-date data for a block is sent to the requestor.

ShimGen classifies obtain transactions as obtaining one of two levels of permission: exclusive read-write permission or shared read-only permission.
This clearly maps to the single-writer, multiple-reader (SWMR) guarantee provided by traditional CPU protocols, but a subset of these same permissions are also present in proposed GPU protocols such as HMG \cite{ren:hpca:2020} and DeNovo \cite{sinclair:micro:2015}.

To determine when a transaction in the cluster directory involves a cache obtaining permission, and which permission it obtains,
ShimGen analyzes the cluster cache controller to determine the permission implied by each cache state, based on two factors:

\begin{itemize}
\item The accesses allowed to hit in the cache in a state, completing without sending a request to the directory, and 
\item Whether the cache downgrades to a state with less permission in response to another cache's read or write, forwarded by the directory.
\end{itemize}

The second condition ensures the analysis detects only permissions visible to the rest of the system.
For example, consistency-directed protocols such as RCC allow a cache to freely read and write to the L1 in the Valid state,
but this state does not have read or write permission, as these accesses are not visible to the rest of the system until a synchronization access.
Detecting this state as having read-write permission would cause ShimGen to obtain exclusive ownership system-wide for any read or write, which is unnecessary to enforce (compound) consistency.

After determining the permission held by each state, denoted by $\perm(s)$, ShimGen can determine which transitions in the directory controller correspond to a cache obtaining new permissions.
If a cache transaction ends in a state with greater permission than it began in---where permissions are ordered greatest-to-least as exclusive read-write,
shared read-only, none---it must have obtained the final state's permission from the directory.
The first message sent by such a transaction in the cache obtains the necessary permission, and directory transactions beginning with this message are classified as upgrades to the given permission level.
Directory transactions obtaining permissions are also classified by whether they send the block's data to the requestor, accommodating features such as Intel's SpecI2M, obtaining permissions without data for full-block writes.

It is not necessary to identify transactions in which a cache downgrades or loses permission, such as an L1 eviction, in our shim API.
For example, when an MSI cluster cache in M issues a PutM, the directory transitions to I without issuing a request in the shim.
From the perspective of the global protocol, the cluster continues to have exclusive ownership and write permission, and will be downgraded only on a read from another cluster.

\subsubsection{One-offs}

The second category of cluster directory transactions classified by ShimGen consists of accesses performed once, at the cluster directory/L2, instead of obtaining permission to be performed later at the requesting cache.
Cluster protocols may include such transactions for consistency-directed protocols employing self-invalidations and flushes rather than obtaining permissions, simpler implementation of atomic accesses, or extending CPU protocols to support streaming/non-temporal accesses.
These one-off access transactions are classified by two parameters: the type of operation performed and its scope.

ShimGen supports three types of one-off accesses: reads, writes, and RMWs.
Each operation is supported in the DSL used to describe protocols to ShimGen, allowing transactions in the directory that perform them to be identified as one-offs.

The other important parameter for one-off accesses is their scope, determining to what domain they must be made visible.
Because these accesses are performed at the cluster directory/L2, they are visible within the cluster as soon as they complete.
However, they do not necessarily need to be made globally visible throughout the system immediately.
Modern CPU and GPU ISAs include a notion of ``scopes" for accesses, allowing differentiation between accesses that must be made visible within a cluster and those that must be made globally visible.

ShimGen allows the user to specify the scope (either cluster or global) of accesses in the cluster protocol.
If a one-off transaction in the cluster directory is caused only by cluster-scoped accesses in the cache controller, it is classified as cluster-scoped.

\subsubsection{Revokes}

The previous two categories of shim transaction correspond to transactions in the cluster directory controller which may require some action in the global protocol.
In the other direction, transactions in the global protocol cache controller may require some action in the cluster protocol.
Because we do not consider update-based global protocols, where accesses from one cluster are performed within another, the only possible consequence of a global cache transaction within the cluster is a change in permission: either downgrading from (exclusive) write to (shared) read permission, or losing permission entirely.
ShimGen uses the same analysis described in Section \ref{sec:identify-obtains} to determine the permission held by each state in the global cache controller.
Each transaction ending in a state with less permission than it began with is classified as a ``revoke," with the resulting permission level.

Table \ref{tab:shim_actions} gives the full list of transaction classifications making up the shim API, along with the classifications ShimGen identifies for the MOESI, CXL, Spandex, and scoped RCC protocols.

\begin{table*}[h]
\scriptsize
\newcommand{\ctg}{Cluster$\rightarrow$Global}
\newcommand{\gtc}{Global$\rightarrow$Cluster}
\centering
\setlength{\extrarowheight}{2pt}
\begin{tabular}{r|l|cccc}
\toprule
 & \textbf{Classification} & \textbf{MOESI} & \textbf{CXL} & \textbf{Spandex} & \textbf{RCC Scoped} \\
\midrule
\multirow{4}{*}{\rotatebox[origin=c]{45}{\makecell{Obtain}}}
 & Shared read permission w/ Data & GetS & RdShared & ReqS &  \\
 & Shared read permission w/o Data &  &  &  &  \\
 & Exclusive write permission w/ Data & GetM & RdOwn & ReqO+data &  \\
 & Exclusive write permission w/o Data &  & RdOwnNoData & ReqO &  \\
\cmidrule(lr){1-6}
\multirow{4}{*}{\rotatebox[origin=c]{45}{\makecell{One-offs}}}
 & Global read &  & RdCurr & ReqV & GetV-Global \\
 & Global write &  & WrInv & ReqWT & PutV-Global \\
 & Cluster read &  & & & GetV-Cluster \\
 & Cluster write &  & WOWrInv & & PutV-Cluster \\
\cmidrule(lr){1-6}
\multirow{2}{*}{\rotatebox[origin=c]{45}{\makecell{Revokes}}}
 & Downgrade to read permission & Fwd-GetS & SnpData & RvkO &  \\
 & Revoke all permissions & Inv/Fwd-GetM & SnpInv & Inv/RvkO &  \\
\bottomrule
\end{tabular}
\caption{Examples of Transaction Classifications in Common Protocols}
\label{tab:shim_actions}

\vspace{-0.25in}
\end{table*}

\begin{algorithm}[t]
\begin{tabular}{@{}p{0.49\linewidth}@{\hspace{0.015\linewidth}\vrule\hspace{0.015\linewidth}}p{0.49\linewidth}@{}}

\begin{minipage}[t]{\linewidth}
\refstepcounter{algorithm}
\label{alg:step1}
\noindent\textbf{Algorithm \thealgorithm.} Classifying protocol transactions\par
\noindent\rule{\linewidth}{0.4pt}\par
\scriptsize

\begin{algorithmic}[1]
\Statex \textit{Inputs: directory controller $D$, cache controller $C$.}

\State $\reqperm, \reqscope \gets \emptyset$

\State \textbf{for each} $\kappa \in \Txns(C)$:
\State \quad \textbf{if} $\perm(\finish(\kappa)) > \perm(\start(\kappa))$:
\State \quad \quad $\reqperm[\firstsend(\kappa, D)] \gets \perm(\finish(\kappa))$
\State \quad $\reqscope[\firstsend(\kappa, D)] \gets \scope(\trigger(\kappa))$

\State \textbf{for each} $\tau \in \Txns(D)$:
\State \quad \textbf{if} $\trigger(\tau) \in \reqperm$:
\State \quad \quad $p \gets \reqperm[\trigger(\tau)]$
\State \quad \quad $\class(\tau) \gets \Obtain(p,\sendsdata(\tau))$
\State \quad \textbf{else if} $\accessesdata(\tau)$:
\State \quad \quad $\class(\tau) \gets \OneOff(\accesstype(\tau),\reqscope[\trigger(\tau)])$

\State \Return $\class$
\end{algorithmic}

\end{minipage}
&

\begin{minipage}[t]{\linewidth}
\refstepcounter{algorithm}
\label{alg:step2}
\noindent\textbf{Algorithm \thealgorithm.} Stronger-than Relation\par
\noindent\rule{\linewidth}{0.4pt}\par
\scriptsize
\begin{algorithmic}[1]

\Statex \textit{$\tau_{GC} =$ global-cache transaction, $\tau_{CD} = $cluster-dir transaction.}

\State $c_{GC} \gets \class(\tau_{GC})$
\State $c_{CD} \gets \class(\tau_{CD})$

\State \textbf{if} $c_{CD}=\Obtain(p,d)$:
\State \quad \Return $c_{GC}=\Obtain(p',d') \wedge p' \ge p \wedge (\neg d \vee d')$

\State \textbf{if} $c_{CD}=\OneOff(\Read,s)$:
\State \quad \Return $(c_{GC}=\OneOff(\Read,s') \wedge s' \ge s)$
\State \quad \quad $\vee\ c_{GC}=\Obtain(\Read,\mathsf{true})$
\State \quad \quad $\vee\ c_{GC}=\Obtain(\ReadWrite,\mathsf{true})$

\State \textbf{if} $c_{CD}=\OneOff(\mathsf{Write},s)$:
\State \quad \Return $(c_{GC}=\OneOff(\mathsf{Write},s') \wedge s' \ge s)$
\State \quad \quad $\vee\ \exists d.\ x=\Obtain(\ReadWrite,d)$

\State \textbf{if} $c_{CD}=\OneOff(\mathsf{RMW},s)$:
\State \quad \Return $(c_{GC}=\OneOff(\mathsf{RMW},s') \wedge s' \ge s)$
\State \quad \quad $\vee\ c_{GC}=\Obtain(\ReadWrite,\mathsf{true})$

\State \Return $\mathsf{false}$
\end{algorithmic}

\end{minipage}
\end{tabular}
\end{algorithm}

\subsection{Step 2: Select Appropriate Transactions}
\label{sec:select-transactions}

After classifying each transaction in the cluster directory and global cache controller by its coherence guarantees relevant for shim generation,
ShimGen must find the appropriate functionality in the other protocol to provide these guarantees.
In some cases, this choice of mapping is clear.
For example, a cluster directory transaction obtaining exclusive write permission and data should map to a global transaction obtaining the same.
However, there are two complications when fusing different protocols.

First, some obtain or one-off transactions in the cluster protocol may not have an analog in the global protocol.
For example, with an RCC cluster connected to an MSI global protocol, a cluster PutV involves a one-off write, but MSI does not have a one-off write.
Instead, this write should obtain exclusive write permission globally and be performed at the cluster's shared L2.

Second, even if the protocol includes the matching access, the architect may want to use a stronger one. For example, when connecting an RCC cluster to a CXL global protocol, a PutV can be mapped to CXL's one-off write (the WrInv message).
However, if writes exhibit locality, it may be better to obtain global SWMR write permission in the cluster L2, using a RdOwn.
We do not take a position on which of these choices should be used, instead allowing the architect using ShimGen to choose between all reasonable and correct options, evaluating the tradeoffs for the system. 

ShimGen solves these issues by introducing a partial order of cluster transaction classifications by ``strength.''
A transaction in the cluster directory can be mapped to any global protocol transaction that provides the same functionality or stronger.
In the example above, obtaining exclusive write permission is considered stronger than performing a one-off write,
so an RCC PutV in a cluster may be made visible globally either by obtaining exclusive write permission or by performing a one-off write at the global LLC.
The details of this strength classification are described next and in Algorithm~\ref{alg:step2}.

Specifically, obtain transactions are ordered by the level of permissions obtained (with exclusive write permission stronger than shared read permission) and whether the cache obtains up-to-date data.
Additionally, transactions obtaining exclusive write permission and data are considered stronger than any one-off access, and those obtaining shared read permission and data are stronger than a one-off read.
As a result, an obtain transaction in the cluster can be mapped to any global transaction obtaining a higher permission level and, if necessary, the data.
A one-off access can be mapped to a matching one-off globally, or it may obtain either read or write permission to be performed in the cluster L2.

When a global protocol supports weakly-ordered accesses (most common in GPU-focused protocols such as HMG~\cite{ren:hpca:2020}), care must be taken to avoid mapping cluster one-off transactions which assume stronger ordering to global accesses which do not provide it.
In this case, ShimGen determines which accesses cause each one-off write transaction to occur, and maps coherence transactions based on user-specified consistency mappings (e.g., from ArMOR~\cite{lustig:isca:2015}).
Note that the shim itself is not responsible for enforcing ordering between accesses, such as flushing the cache on a release to ensure it is ordered after previous writes.
Rather, ShimGen maps each one-off access to the appropriate transaction in the global protocol, which is assumed to correctly implement the ordering guarantees implied by its consistency model.

For each transaction at the cluster directory classified in step 1, ShimGen finds all global protocol transactions that provide the same or stronger functionality.
By default, ShimGen maps the cluster transaction to the weakest of these global transactions, providing the most similar semantics.
However, ShimGen also lists the possible stronger transactions and allows the user to override this default to evaluate alternative possibilities for protocol design.

ShimGen must also identify relevant transactions in the other direction, mapping global cache downgrades to cluster transactions which downgrade all caches in the cluster.
To do this, ShimGen uses the ``proxy cache'' mechanism discussed in prior work~\cite{oswald:hpca:2022}.
When the cluster loses all global read and write permissions for a block, due to another cluster obtaining exclusive ownership or the cluster's directory evicting the block,
ShimGen initiates the transaction in the cluster directory that would be caused by an SWMR write request from within the cluster.
When a forwarded read request requires downgrading any exclusive owner to shared read permission, ShimGen initiates the transaction that would be caused by an SWMR read request coming from the cluster.
If no request obtaining shared permission exists in the cluster (e.g., for a DeNovo cluster including only an exclusive Registered state), the stronger transaction obtaining exclusive permission is used for both.
If no transactions obtain shared or exclusive permission in the cluster, as in a GPU protocol using self-invalidated, write-through L1s, no action needs to be taken to downgrade the cluster.

Triggering these transactions at the cluster directory requires adding some additional state to track the status of the proxy cache transaction.
The necessary states and transitions are automatically generated by ShimGen, following the same approach as prior work, and are added to the directory controller finite state machine for the purposes of ShimGen's algorithm.

\subsection{Step 3: Combining Transactions}

With the previous two steps, ShimGen identifies a mapping from transactions occuring in the cluster directory to those which must be performed in the global cache to maintain coherence, and vice versa.
In the final step, ShimGen determines how to correctly interleave the operations comprising these transactions when the controllers are fused with a shim.

If every transaction obtains permission and data, there is a straightforwardly correct---but not necessarily optimal---approach to this fusion.
When a cluster directory transaction requires global SWMR permission, \emph{encapsulate} the global transaction within the cluster directory transaction, deferring the incoming request from the cluster until the global protocol transaction has completed and permissions have been obtained.
This ensures that a cache within a cluster cannot obtain permission not yet held in the global protocol.

However, this approach is not correct when generalizing beyond SWMR global protocols, and in other cases introduces unnecessary serialization, increasing latency.
ShimGen uses a more sophisticated approach, considering the precise coherence guarantees required by each type of transaction classified in step 1.
Rather than performing the entire global transaction, followed by the cluster transaction, ShimGen allows both to proceed in parallel while enforcing dependencies between them required to maintain coherence/consistency.
The following subsections describe the dependencies required for each type of transaction; the details of how dependency-respecting parallel composition is implemented are described in Section~\ref{sec:step3compdetails}.

\subsubsection{Obtains}

When a cluster transaction obtains permission and/or data, its interaction with the global protocol has two requirements: the requesting cluster cache should not upgrade its permission until permission is obtained globally, and (if the transaction obtains data) data cannot be sent to the requestor until the most up-to-date data is obtained in the global protocol.

Both of these constraints are satisfied by waiting for the global protocol transaction to complete before proceeding with any of the cluster transaction, but this is not necessary.
For example, invalidating sharers within a cluster can proceed before or in parallel with obtaining exclusive ownership globally, and data can be sent to a requestor when it is available from the global protocol, even if permission has not yet been acknowledged.

In general, ShimGen analyzes upgrade transactions within the cluster to determine which operations can be performed independently of the global transaction, which require only the data from the global transaction, and which must wait until global permission is obtained.
First, ShimGen analyzes the L1 cache controller transaction that causes the upgrade, identifying the final message received from the directory before permission is obtained.
This message cannot be sent by the directory until global permission is obtained.
Second, ShimGen finds any operations in the cluster directory which require up-to-date data from the global protocol (e.g., sending data to the requestor).
Such operations can be performed after data is obtained from the global protocol, even before the transaction is complete.
For protocols, including CXL, which separate acknowledgments and data messages, this avoids unnecessary serialization at the directory.

Operations not in one of the above two categories can proceed in parallel with the global transaction, with one subtlety:
a cluster operation can only be performed once it is known to be necessary, regardless of any other message received during the transaction.
For example, a 3-hop MSI protocol upgrading from S to M normally sends invalidations to sharers within the cluster, which forward acknowledgments to the requestor.
However, if an invalidation message is received from the global protocol, invalidated sharers must instead forward acknowledgments to the directory, which completes the global invalidation before upgrading from I to M.
If invalidations were sent to sharers in parallel with the global upgrade, they would incorrectly forward acknowledgments to the requestor.
To avoid this,
ShimGen explores all possible traces of transitions which can occur in the global protocol prior to the obtain transaction completing from each state. Only once a cluster action will occur on all subsequent global traces can it be performed.

\subsubsection{One-offs}

A different approach is needed for generating one-off transactions, which perform an access in a lower-level cache rather than obtaining permission to perform it in the L1.
In particular, waiting for the entire global transaction to complete before handling the initial request at the cluster directory, as in prior work, is not only suboptimal, but impossible.
One-off accesses typically require data to be sent from the requesting L1 to the cluster directory/L2, often separate from the original request.
For example, in a CXL Write transaction, the cluster directory must send a WritePull message to the requestor to obtain the data to be written before performing a write.

Because of this, ShimGen interleaves the operations of a cluster one-off transaction and the associated global transaction to satisfy two constraints:
(1) data cannot be used by the global protocol until it is received in the cluster directory (and vice-versa), and (2) for global-scoped transactions, the final acknowledgment within the cluster should not be sent until the global transaction is complete to ensure consistency.
Other than this, the two transactions can proceed concurrently, subject to the same limitation discussed above, whereby cluster operations cannot be performed until they are known to be necessary regardless of the global transaction.

For example, if a global protocol and a cluster protocol each use the same CXL Write transaction format, ShimGen would identify the following interleaving of operations:
\begin{itemize}
    \item When the cluster Write request is received, issue a global Write request, a cluster WritePull to get data from the requestor, and Invalidations to any sharers.
    \item When both the WritePull from the global directory and Data from the requestor have arrived, send the Data globally.
    \item When both the final invalidation acknowledgment from the cluster and the Global Observation (GO) acknowledgment from the global protocol have arrived, send the GO acknowledgment in the cluster.
\end{itemize}
ShimGen's algorithm avoids unnecessary serialization of operations within the cluster and global protocol---for example, waiting for the global WritePull to arrive before calling back data from the requestor or invalidating sharers---while ensuring the correctness of the transaction flow.

For cluster-scoped operations, ShimGen does not require an access to complete globally before proceeding with the cluster transaction.
For example, an RCC PutV sent on a cluster-scoped release can be acknowledged as soon as it is received at the cluster directory.
To ensure consistency, later global-scoped transactions must wait for any in-flight cluster-scoped transactions to complete globally before proceeding.
ShimGen includes the necessary actions to track outstanding cluster-scoped transactions and wait for them to complete before executing a global-scoped transaction in the generated shim.
The user of ShimGen wishing to avoid this overhead may disable this optimization for cluster-scoped operations.

\subsubsection{Revokes}

Interleaving a global cache revocation transaction with a corresponding proxy cache transaction is simpler than the cache upgrades described previously.
Typically, such a transaction only sends a single message in the global protocol, acknowledging that permissions have been revoked.
Such a message cannot be sent until the invalidation transaction is complete in the cluster.

However, care is needed to handle revocation transactions concurrent with upgrades from the cluster.
Because revocations are initiated by the global protocol, they 
serialize before any ongoing upgrade from the cluster.
As such, for every transient state in which the cluster directory waits for a global upgrade or one-off to be acknowledged, ShimGen inserts transitions to handle revokes from the global protocol, as if they had been received before the upgrade was initiated.

For example, an MSI cluster directory in state S receiving a GetM from one its L1 caches requests global exclusive write permission and transitions to transient state SM$^\text{G}$.
From this state, ShimGen adds transitions to handle a revocation of all permissions, as if it had been received in the original stable state, S.
This transaction, generated with the proxy cache mechanism as described in Section \ref{sec:select-transactions}, invalidates all sharers before acknowledging the global revoke.
When it completes, it transitions to state IM$^\text{G}$, as if the original GetM had arrived in state I, after the global transaction.

\newcommand{\dep}{\mathsf{dep}}
\subsubsection{Generating Fused Transactions}
\label{sec:step3compdetails}

The logic above determines which operations in each controller's transaction must wait for certain events to have occurred in the other controller.
ShimGen updates the graph of each transaction to encode this information via a new type of guard, $\dep(D_e)$, indicating that a transition cannot occur until the other controller has reached some state in set $D_e$.

When a transition $A\xrightarrow{g/a}B$ includes operations which require some event $e$ to have occurred in the other controller, the transition is split into two, $A\xrightarrow{g/a'}A'\xrightarrow{\dep(D_e)/a''}B$, where $D_e$ is the set of states in the other controller in which the dependent operations $a''$ can occur.
For example, an MSI cluster directory needing to wait for the global protocol to enter state $D_e = {M}$ before acknowledging a GetM would be represented by the graph $I \xrightarrow{\mathrm{GetM}/} IM^G \xrightarrow{\dep(D_e)/\mathrm{send\,ack}} M$.

Once dependency information is added to each protocol's transaction graph, ShimGen generates a single fused transaction which allows both transactions to run in parallel while respecting dependency requirements.
Specifically, when the cluster directory is in state $c$ and global directory in state $d$, any transition $c \to c'$ or $g \to g'$ without a $\dep$ guard is replaced with a transition $c \times g \to c' \times g$ or $c \times g \to c \times g'$, respectively.
If the resulting state would allow for a transition with guard $\dep(D_e)$ with one of $c,c',g,g' \in D_e$, signifying that operations previously blocked on another controller's action are now possible, this transition is chained onto the end of the generated transition so the associated operations are performed immediately.

\subsection{ShimGen Output}

With the above three steps, ShimGen generates the shim logic to combine a cluster protocol directory with a global protocol cache controller.
By generating a shim for every cluster protocol, then combining them with a global protocol, ShimGen generates a complete hierarchical protocol.
ShimGen can output this protocol in the same input DSL, the Murphi model checking language for verification, and the SLICC DSL for benchmarking with gem5.

%% file: case-studies.tex
\section{Case Study 1: APU Protocol}
\label{case}

In this section, we compare the protocol output by ShimGen to an existing, manually developed hierarchical protocol.
The protocol in question was released with gem5 as part of the AMD APU simulator. A global MOESI protocol connects a CPU cluster with an MESI protocol and a GPU cluster with a write-through release-consistency-based protocol.
Using ShimGen, we model the cluster and global protocols separately in ShimGen's input DSL and compare the shims generated by ShimGen to the original gem5 implementation.
We now describe each protocol, how it is analyzed in ShimGen, and how the shims generated by ShimGen for each cluster compare to the manually-designed protocol.

\subsection{Global MOESI Protocol}

The global MOESI protocol, unlike typical directory protocols, uses a stateless directory.  Rather than tracking sharers and owners, the directory forwards downgrades/invalidations to all clusters on a read/write request, collecting responses before acknowledging the request.
The protocol also includes a WriteThrough transaction, which invalidates sharers and performs a write directly to the shared LLC, rather than granting ownership of a block to a cluster.\footnote{The global MOESI protocol includes coarse-grained region coherence~\cite{region-coherence}, which we do not model in ShimGen.}

Despite the stateless directory, ShimGen identifies SWMR upgrades and downgrades the same as a standard directory-based MOESI protocol.
ShimGen also identifies the WriteThrough transaction as a one-off write.

\subsection{CPU MESI Protocol}

The APU simulator organizes CPUs into ``Core-Pairs'' comprising two CPU cores with separate L1 data caches sharing an L2 cache, kept coherent with a MESI protocol.
While the gem5 protocol uses a single controller for both the L1 and L2, we translate this model to the fully-specified protocol with separate L1 and directory/L2 controllers that ShimGen expects, adding the necessary messages between L1 and L2.

Aside from this modeling difference, interactions between the CPU MESI and global MOESI protocol behave identically in the ShimGen-generated protocol and the manually-designed one.
GetS and GetM requests sent on an L1 load or store miss are identified by ShimGen as SWMR reads and writes, and translated to RdBlk and RdBlkM transactions in the global protocol.
In the other direction, invalidation and downgrade probes from the global protocol are mapped to the corresponding invalidation/downgrade in the cluster.

\noindent{\textbf{Summary}: ShimGen is equivalent to manual protocol.}

\subsection{GPU VIPER Protocol}

The GPU cluster uses VIPER, a write-through protocol that enforces RC by invalidating the L1 on an acquire and waiting for all prior write-throughs to be acknowledged on a release.
Unlike the CPU protocol, the gem5 GPU protocol is modeled as separate L1 and L2 controllers, with the L2 translating messages between the cluster and global protocol.
We remove interaction with the global protocol from the L2 controller to model a single-level VIPER protocol, which we can then fuse with the global protocol using ShimGen.

Because GPU caches do not obtain read or write permissions for blocks, forwarding subsequent accesses to owners/sharers, ShimGen classifies reads and write-throughs received at the L2 in the VIPER protocol as one-off reads and writes.
The enforcement of RC by invalidating the L1 on an acquire and waiting for prior write-throughs to serialize on a release is implemented solely within the L1 controller, and thus is not relevant for generating shims.

When fused with the global MOESI protocol, either using ShimGen or in the manually-designed gem5 L2 controller,
VIPER reads that miss in the L2 issue a read request in the global MOESI protocol, downgrading any writer and obtaining up-to-date data before replying to the requester.
There is one difference between the handling of concurrent reads to the same block in the two controllers.
When waiting for a response from the global protocol, ShimGen's controller stalls any subsequent reads to the same block from other L1s, responding to all when the global protocol response is received.
The manually designed protocol instead sends an additional request in the global protocol if another read to the same block arrives from the cluster.
Each global request and response specifies the original requesting L1, allowing a read to be acknowledged by the L1 when the corresponding global request is acknowledged.

VIPER write-throughs, classified by ShimGen as one-off writes, can be translated to the global protocol in two different ways.
First, the global protocol WriteThrough transaction may be utilized, writing through the L2 rather than reading the block for ownership.
Both ShimGen and the manually-designed L2 controller implement this option similarly, issuing a WriteThrough to the global protocol and responding to the original request once it is acknowledged.
As with reads, ShimGen allows only one in-flight write to a given block, while the manually-designed protocol can issue multiple WriteThroughs simultaneously for a single block.

The only difference between the gem5 and ShimGen protocols relates to support for a write-back L2, in which 
writes are performed only in the L2 and are written back globally only on an eviction or read from another cluster.
ShimGen supports this by allowing the one-off write to be mapped to a RdBlkM transaction in the global protocol, invalidating sharers globally before transitioning to M and acknowledging the write.

\begin{figure}
    \centering
    \includegraphics[width=0.55\linewidth]{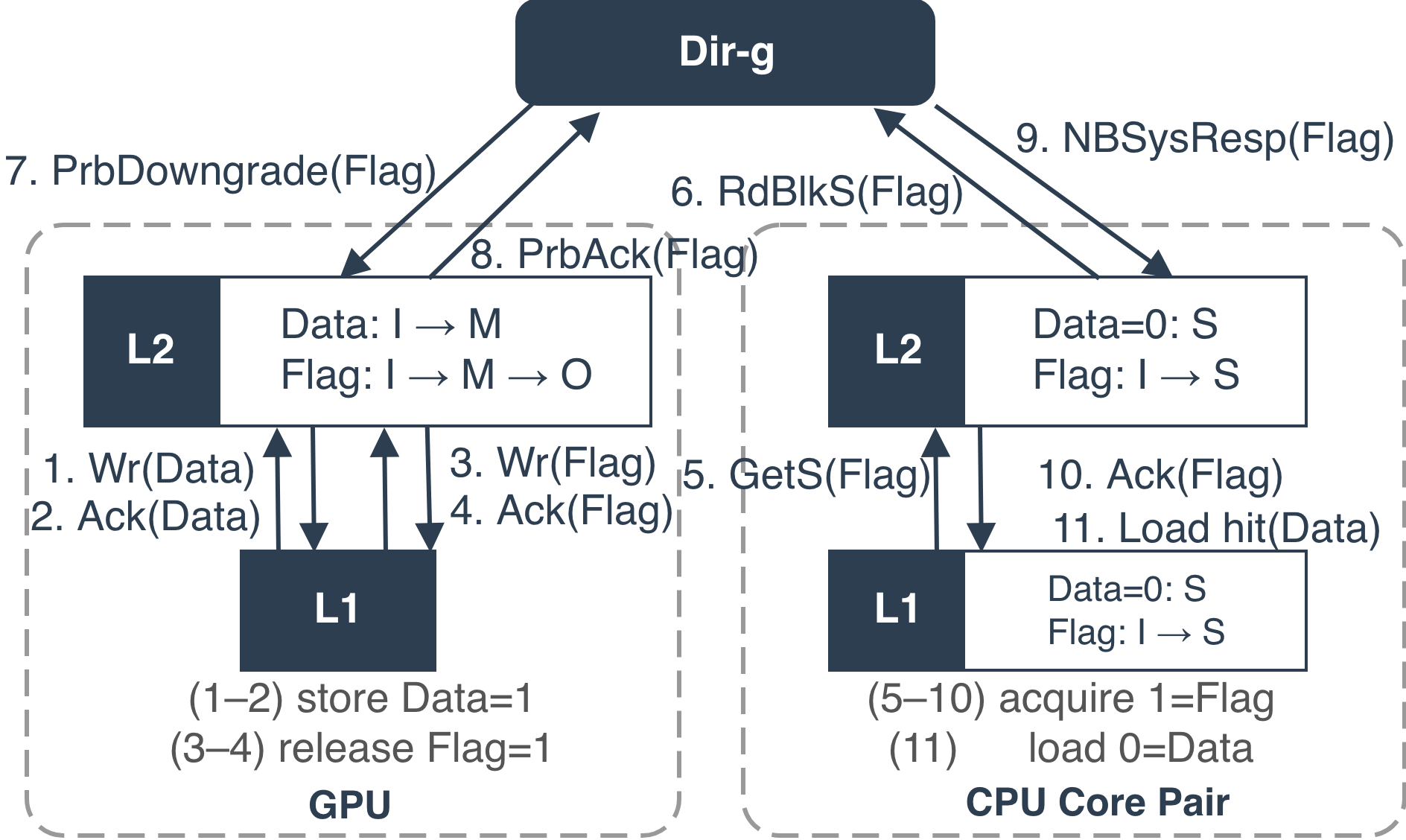}
    \caption{Message Passing litmus test on APU Protocol}
    \label{fig:mp-gem5}
\vspace{-0.15in}
\end{figure}

The manually-designed L2 controller instead transitions directly to M on a write in write-back mode without obtaining permission globally.
However, this design fails to enforce release consistency, as exemplified by the message-passing litmus test between GPU and CPU threads shown in Figure \ref{fig:mp-gem5}.
If a CPU has the block containing Data cached in state S when it is written by the GPU, the GPU will write the block to its L2 without invalidating the CPU.
Even after a subsequent release of a Flag by the GPU and CPU acquire seeing the new value, the CPU is left with stale data in its cache.
Thus, the outcome Flag=1, Data=0 forbidden by RC is observable.

The controller generated by ShimGen avoids this RC violation by ensuring writes to the GPU L2 invalidate any CPU cache in S.
However, this invalidation adds the latency of the global protocol to all GPU writes to blocks not already held in M in the L2, something which the designers of the protocol may have been trying to avoid.
ShimGen can avoid this latency penalty while enforcing RC with a slight modification to the VIPER protocol.
If write-through requests from the L1 are tagged as either unordered writes or globally visible releases, ShimGen classifies the former as cluster-scoped writes, rather than assuming both must be made globally visible.
This allows the L2 to respond to non-release writes immediately, while still enforcing consistency by acknowledging releases only when all prior writes are made globally visible.

Although it is unclear if the designers intended to support compound consistency, we have found that the manual protocol fails to uphold it. This case study illustrates that manually optimizing hierarchical protocols often risks correctness, whereas ShimGen provides a systematic path to synthesize correct protocols.

While prior work on synthesizing CXL bridges~\cite{lefort:asplos:2026} could generate a similar protocol, including avoiding the consistency violation, it is not compatible with the custom stateless-directory MOESI protocol used by the AMD APU.
Prior work also assumes each write in the cluster requires obtaining global SWMR permission before acknowledging each write, while ShimGen additionally supports the use of global WriteThroughs or early acknowledgments to avoid the latency penalty for non-release writes.

\noindent{\textbf{Summary}: ShimGen is mostly equivalent to the manual protocol except in one aspect, in which ShimGen's protocol enforces consistency and the manual protocol does not.  ShimGen does so while allowing similar optimizations as in manually-written protocols.
}

\subsection{Performance Evaluation}

To evaluate the overhead of ShimGen's synthesized shims,
as well as the importance of non-SWMR global protocol behavior,
we compare the performance of four different configurations of the APU protocol using gem5: the manually designed APU protocol in both write-back and write-through modes, ShimGen's protocol generated with GPU write-throughs obtaining global SWMR permission, and ShimGen's protocol making use of global one-off writes.
All experiments use the AMD APU syscall emulation (SE) mode sample configuration included with gem5, modeling a system integrating a gfx902 GPU cluster and 4-core CPU cluster.
The latter two configurations modify only the GPU L2 (TCC) controller state machine to reflect the differences described above.

We run ten GPU benchmarks on each of four protocols: PENNANT~\cite{ferenbaugh:pennant}, three graph workloads from the Pannotia~\cite{che:pannotia} suite, and six synchronization microbenchmarks from HeteroSync~\cite{sinclair:heterosync}.
Across all runs displayed in Figure~\ref{fig:apu-bench}, the synthesized protocol runtimes ranged from 0.90\% slower to 1.04\% faster than the manually-designed controller, with an average 0.13\% performance improvement.
Thus, ShimGen introduces little to no overhead compared to the manually-designed protocol.

Between write-through and write-back modes, most benchmarks also showed little difference.
However, one workload, the PENNANT mesh-based physics simulator, showed an 18\% decrease in runtime when one-off writes were used (for both the manual and synthesized variants).
This shows the importance of supporting non-SWMR global protocol features for some workloads.

\section{Case Study 2: Non-Temporal Streaming Writes}
\label{sec:perf-case-study}

Our second case study demonstrates the performance implications of the different coherence functionality supported by ShimGen.
We consider a system of several CPU clusters connected together with a CXL.cache global protocol in a multi-host configuration.
Each cluster uses an MSI protocol internally, while also including a CXL.cache directory managing inter-cluster coherence for the memory backed by that cluster.
Using ShimGen, we generate cluster LLC controllers combining a cluster-facing MSI directory controller and global-facing CXL cache controller.

In this case study, we focus on non-temporal stores, a common feature of CPU ISAs hinting that a store does not exhibit locality.
For such stores, reading the accessed block into the cache wastes bandwidth over performing the store directly to memory.
A protocol that avoids these unnecessary read-for-ownership reduces coherence traffic for streaming workloads using these accesses.

To evaluate how this optimization impacts ShimGen's hierarchical protocols, we extend our MSI protocol with a non-temporal store transaction performed at the LLC.
A non-temporal store to a cache in I or S is sent to the directory, which invalidates sharers and performs it in the LLC, before acknowledging the access.

As a baseline, we generate shims which obtain SWMR read or write permission in the global protocol for every MSI protocol transaction, including non-temporal stores.
This matches the behavior of prior work on the synthesis of hierarchical coherence protocols.
However, ShimGen can also generate shims which make use of similar one-off transactions in the global protocol.
For our CXL.cache global protocol, a non-temporal store transaction can be mapped to a WrInv.
We use ShimGen to generate both the baseline read-for-ownership protocol and the protocol making use of one-off writes, and output SLICC code for benchmarking in gem5. 

To compare the performance of these two methods for handling non-temporal stores, we modify the STREAM microbenchmark~\cite{mccalpin:1995} to annotate stores as non-temporal.
This microbenchmark copies a large array of data from one location in memory to another, measuring the effective memory bandwidth of the system.
By varying the stride of the array accesses, we can decrease the number of accesses per cache block, increasing the benefit of non-temporal accesses to avoid unnecessarily reading or writing entire cache blocks.

\begin{table}[]
\small
    \centering
\begin{tabular}{r|l}
        Cores & x86 cores, 8-wide OoO, 128 KB private L1, 1 cycle controller latency \\
        \hline
        Clusters & 4 clusters of 4 cores each, 4 MB shared L2 per cluster \\
        \hline
        Global Interconnect & mesh topology, 70 ns link latency, 1 clk router latency, 72B flit \\
        \hline
        Memory & DDR5, 4400 Mhz, 8GB, 10ns controller \\
    \end{tabular}
    \caption{System Parameters for Case Study 2}
    \label{tab:case-study-system-params}
    \vspace{-0.20in}
\end{table}

\begin{figure}[t]
    \centering
    \begin{minipage}[t]{0.56\linewidth}
        \centering
        \includegraphics[width=\linewidth]{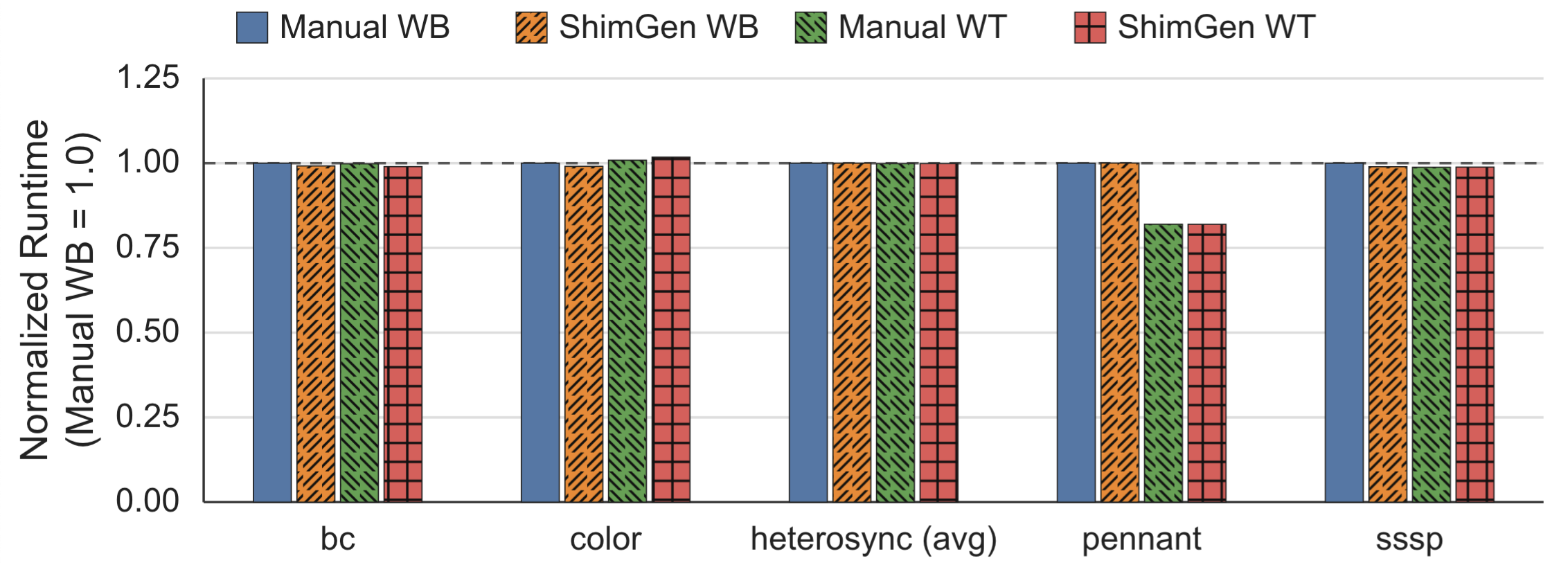}
        \caption{Performance of Manually-Designed and Synthesized APU Protocol in WB (SWMR) and WT (One-off) Mode}
        \label{fig:apu-bench}
        \Description[Normalized runtimes for four APU protocol variants]{Grouped bar chart of normalized runtime for bc, color, heterosync average, pennant, and sssp. Manual and ShimGen write-back and write-through variants are near the baseline for most workloads; pennant is substantially faster with write-through variants.}
    \end{minipage}\hfill
    \begin{minipage}[t]{0.40\linewidth}
        \centering
        \includegraphics[width=\linewidth]{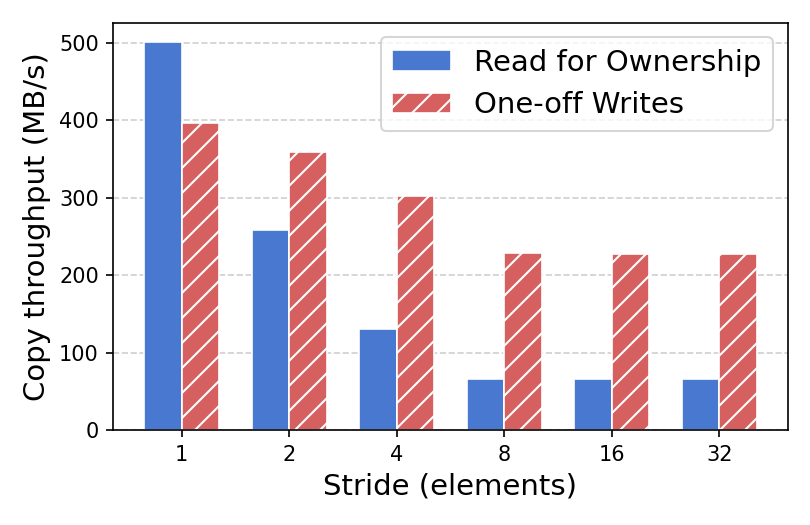}
        \caption{STREAM Copy Throughput with Read-for-Ownership and One-off Mapping}
        \label{fig:stream-results}
        \Description[STREAM throughput by access stride]{Bar chart comparing STREAM copy throughput for read-for-ownership and one-off writes at strides from 1 to 32 elements. Read-for-ownership throughput falls sharply as stride increases, while one-off writes provide higher throughput for strides of 2 and above.}
    \end{minipage}
\end{figure}

The system parameters for our simulated system are shown in Table \ref{tab:case-study-system-params}, and the results for each protocol are shown in Figure \ref{fig:stream-results}.
Mapping non-temporal stores to a read-for-ownership requires each access to load an entire cache block from the global protocol, decreasing effective memory bandwidth as stride increases and most of the block goes unused.
Respecting the non-temporal hint and performing one-off WrInv transactions in the global protocol avoids any unnecessary network traffic and increases bandwidth for larger strides.
Neither the cluster nor global protocol implement any coalescing of one-off writes, so small strides performing more one-off writes per block require more accesses to main memory than the read-for-ownership protocol.
However, even at a stride size of 2, the decreased network traffic outweighs the increased number of DRAM accesses.

With this case study, we do not aim to show that protocols generated by ShimGen have significantly higher performance on general workloads.
However, the existence of features such as one-off writes in both academic proposals~\cite{alsop:isca:2018} and industrial standards~\cite{cxl, chi} indicates their importance for some workloads.
As this case study demonstrates,
ShimGen's flexibility enables architects to exploit these and other coherence protocol transactions and optimizations.

%% file: verification-short.tex
\section{Verification}
\label{sec:verification}

We have shown how ShimGen can synthesize hierarchical protocols for a variety of systems.
However, it is also essential to verify that the resulting protocols are correct.
To do this, ShimGen includes a backend that outputs hierarchical protocols in the Mur$\phi$~\cite{dill:cav:1996} model checking language.

To evaluate the correctness of ShimGen's algorithm in general, we generate Mur$\phi$ models for a wide variety of protocols.
We consider all possible systems comprising up to three clusters using MSI (including Section~\ref{sec:perf-case-study}'s variant with non-temporal stores), MOESI, RCC, and CHI protocols, connected with MSI, MOESI, CXL, CHI, or Spandex global protocols.
We also model a multi-GPU system using RCC for both the cluster and global protocol.

For each system, we first use bounded model checking to verify liveness, ensuring the protocols generated by ShimGen are deadlock-free.
We then additionally verify that each hierarchical protocol correctly enforces its memory consistency model.
If each cluster protocol enforces the same MCM, the hierarchical protocol should enforce that same MCM.
In a heterogeneous system in which different cluster protocol enforce distinct consistency semantics, correctness is defined by the recently proposed compound consistency~\cite{DBLP:journals/pacmpl/Goens0SAON23}; this is not a new MCM but is rather a compositional amalgamation of the original MCMs in which threads belonging to each cluster continue to adhere to their original MCMs.

The state of the art in verifying that protocols enforce MCMs is the use of litmus tests: small, multithreaded programs crafted to test critical behaviors of the MCM~\cite{mador-haim:cav:2010, alglave:lncs:2011, manerkar:15}. 
For each system, we generate variants of 7 common litmus tests—MP, Dekkers, IRIW, RWC, WRC, ISA2, and CoWR—considering all mappings of threads to clusters and choices of consistency labels of accesses.
We use the LOST-POP~\cite{DBLP:journals/pacmpl/Goens0SAON23} formalization of compound consistency to determine which litmus test variant outcomes are forbidden by the system's MCM; observing such an outcome indicates a protocol bug.
In total, we generate between 52 and 444 litmus test variations for each system, depending on the MCMs of its protocols.

For each protocol and litmus test variation, we use exhaustive model checking with the Mur$\phi$ model checker to explore all possible execution traces, including all possible interleavings of accesses and protocol messages, as well as arbitrary insertions of cache replacements and non-binding prefetches.
The resulting Mur$\phi$ models are very large, due to the need to model all controller and network state throughout the entire system, and the reachable state space exhausts over 200 GB of memory without completing.

To enable exhaustive model checking for these large models, we use the common technique of abstraction refinement~\cite{DBLP:conf/popl/CousotC77,clarke:92,mcmillan:01,clarke:89}.
For each cluster and global protocol making up a system, we develop an abstraction which preserves the behavior of the original protocol and shim with far less state.
While the abstraction must reproduce all observable behaviors of the concrete protocol---including, notably, all possible concurrent transactions and race conditions occurring in transient states---many messages internal to the protocol and their associated state can be removed.
This enables model checking to be broken into two steps.
First, we verify with Mur$\phi$ that each protocol in a system refines its abstraction: all behavior of each protocol is allowed by its abstraction.
Second, we compose these abstractions to create a hierarchical protocol, which must likewise retain all behavior of the generated hierarchical protocol being tested.
We then run litmus test variants on this abstracted system, and verify that no executions violate compound consistency.
Each of these steps involves model checking a smaller system, avoiding the state space explosion problem of model checking the complete system.
Because all executions allowed in the generated protocol are also possible in the abstracted system, this ensures the generated protocol enforces compound consistency for these litmus tests.
Formally, this follows from two well-known results from model checking: that simulation relations factor over this handshaking composition and preserve safety properties (cf. Theorem 7.4 of~\cite{DBLP:conf/tcs/Park81} and Corollary 7.68 of~\cite{DBLP:books/daglib/0020348}, respectively).

This compositional abstraction technique allows us to exhaustively check all litmus tests in Mur$\phi$, without resorting to bounded model checking.
In all litmus test variations, our constructed protocols respect their compound MCMs, never violating safety.

%% file: related-work.tex
\begin{table*}[hbt!]
\small
\centering
\setlength{\extrarowheight}{2pt}
\begin{tabular}{r|ccccc}
System & Hierarchy & Heterogeneity & Scopes & Global Protocol & Automated Synthesis \\
\midrule
ProtoGen~\cite{oswald:isca:2018} &  &  &  & None & \checkmark \\
HieraGen~\cite{oswald:isca:2020} & \checkmark &  &  & SWMR & \checkmark \\
HeteroGen~\cite{oswald:hpca:2022} &  & \checkmark &  & None & \checkmark \\
$C^3$~\cite{lefort:hpca:2026} & \checkmark & \checkmark &  & SWMR subset of CXL &  \\
vCXLGen~\cite{lefort:asplos:2026} & \checkmark & \checkmark &  & SWMR subset of CXL & \checkmark \\
ShimGen & \checkmark & \checkmark & \checkmark & Choice & \checkmark \\
\end{tabular}
\caption{Comparison of ShimGen and Related Work}
\label{tab:related-work}
\end{table*}

\section{Related Work}
\label{sec:relwork}

\noindent\textbf{System Construction.}
Fractal coherence~\cite{zhang:micro:2010}, Neo protocols~\cite{matthews:micro:2017}, and Hierarchical Cache Coherence (HCC)~\cite{ladan-mozes:spaa:2008} present hierarchical protocols with regular structures that facilitate verification.
None of the prior works considered scoped or non-SWMR protocols. \emph{Our work is the first to consider HHS protocols where the cluster and global protocols can be SWMR, non-SWMR, scoped, or unscoped.} 

Prior work has studied heterogeneous system design. 
Crossing Guard~\cite{olson:asplos:2017} presents a methodology for designing shims that a global protocol can use to safely interface with 3rd-party cluster protocols. However, they only consider SWMR protocols.  $C^3$~\cite{lefort:hpca:2026} presents a generic shim design, with the limitation that the global protocol is the SWMR subset of CXL. 
Alsop et al.~\cite{alsop:isca:2018} present a  global directory controller, Spandex, that embraces features of SWMR and non-SWMR protocols. 
However, Spandex does not consider scoped protocols, and their evaluation does not consider hierarchy---there is no global protocol.  While Spandex's interface supports some non-SWMR features, it does not support RC reads and writes or separate acknowledgments and data.
MemGlue~\cite{cleaveland:fmcad:2024} is a global protocol that supports the C11 MCM and uses shims to interface between MemGlue and cluster protocols. \emph{In contrast to each of the above works, we are the first to consider scoped protocols and the combination of hierarchy and heterogeneity.}

\noindent{\bf Protocol Synthesis}
Recently, vCXLGen~\cite{lefort:asplos:2026} synthesized shims to interface cluster protocols to a global protocol that is the SWMR subset of CXL, as shown in Figure~\ref{fig:mappings-comparison}(top). Unlike ShimGen, vCXLGen cannot accommodate the interest in non-SWMR features in global protocols and their performance potential, nor can it accommodate SWMR protocols besides the subset in CXL.  
HeteroGen~\cite{oswald:hpca:2022} automatically synthesizes heterogeneous protocols by fusing them at a single shared directory, i.e., there is no hierarchy or global protocol. 
Unlike ShimGen, HeteroGen's system model has no global protocol.
Thus, an architect who wants to compose cluster protocols with, say, a global CXL or CHI protocol cannot use HeteroGen for this. 
HieraGen~\cite{oswald:isca:2020} synthesizes hierarchical protocols, but all protocols must enforce SWMR. 
Thus, HieraGen does not support fusing a CPU protocol with a GPU protocol, since GPU protocols do not enforce SWMR.
Neither HieraGen nor HeteroGen handles scoped  protocols, nor can either generate the hierarchical, heterogeneous protocol for a system like  GraceHopper which integrates an SWMR CPU protocol with a scoped, non-SWMR GPU protocol.

We summarize related work compared to ShimGen in Table \ref{tab:related-work}.

%% file: conclusions.tex
\section{Conclusions}

We have developed a tool, ShimGen,  that 
automatically synthesizes a complete hierarchical protocol from a heterogeneous set of cluster protocols and global protocol.  
Unlike prior work, ShimGen supports a wide range of global protocols, including protocols with non-SWMR features.  Through two case studies, we show both that the synthesis of these protocols can lead to subtle violations of compound consistency and that the incorporation of non-SWMR features in global protocols can offer performance benefits.